# Assembly and speed in ion exchange based modular phoretic micro-swimmers

**Ran Niu\*, Denis Botin, Julian Weber, Alexander Reinmüller, Thomas Palberg**
Institut für Physik, Johannes Gutenberg-Universtät Mainz, Staudingerweg 7, 55128 Mainz, Germany

**Abstract**

We report an experimental study on ion-exchange based modular micro-swimmers in low-salt water. Cationic ion-exchange particles and passive cargo particles assemble into self-propelling complexes, showing self-propulsion at speeds of several microns per second over extended distances and times. We quantify the assembly and speed of the complexes for different combinations of ion exchange particles and cargo particles, substrate types, salt types and concentrations, and cell geometries. Irrespective of experimental boundary conditions, we observe a regular development of the assembly shape with increasing number of cargo. Moreover, the swimming speed increases stepwise upon increasing the number of cargo and then saturates at a maximum speed, indicating an active role of cargo in modular swimming. We propose a geometric model of self-assembly to describe the experimental observations in a qualitative way. Our study also provides some constraints for future theoretical modelling and simulation.

**Introduction**

The past 50 years witnessed the appearance and development of micro and nano machines, inspired by biology and guided by theoretical modelling and simulation.[1-10] Overcoming the perturbation from Brownian motion and viscous force is possible by exploiting non-reciprocally moving components (flagella or cilia) or a gradient of some kind combined with a responsive surface leading to phoretic slip between solvent and surface. In line with this principle, a variety of artificial swimmers have been designed, including catalytic particles,[11,12,13] thermo-phoretic nano-particles,[14,15] micro-bubble rockets,[16,17] magnetically driven swimmers,[18,19] and swimmers based on thermo-elastic deformation.[20] With few exceptions, swimmers are formed by single active colloids (or bacteria) which combine all relevant functions in one and the same entity. This facilitates their treatment by advanced statistical mechanics and detailed studies of their collective properties from clustering and swarming to living crystals and meso-scale dynamic patterns.[21]

In some cases, however, several sub-units of different functionalities are combined. For example, a thermo-phoretic gold or Janus particle with an attached DNA-origami thread shows a rectified motion[22] and a catalytic platinum sphere connected with a silica particle forms a self-propelling catalytic dimer.[23] Already early, this modular approach has been studied theoretically e.g., for a hot solid particle propelled by a gas bubble under thermo-capillary flow,[24,25] even though this model can be realized experimentally only under microgravity environment due to buoyancy effects. More recently, this approach was taken in theoretical works combining catalytically active with catalytically inert spheres to form dimer motors.[26,27]

We here realize the modular approach experimentally by assembling swimmers from individual components, which do *not* show any active swimming by themselves.[28] The modular construction allows for a high level of task sharing and promises an optimized swimming performance by tuning individual components. To be specific, the central part of our modular swimmer is a mobile electro-osmotic (eo) pump. The pump is based on a spherical ion exchange resin particle (denoted as IEX$x$ with $x$ representing the diameter), which exchanges cationic impurities for stored protons.[29] Different diffusion coefficients of the exchanged cations and protons generate a local diffusio-electric field pointing inward.[30,31] This field induces a converging electro-osmotic solvent flow on a negatively charged substrate. The physics of this flow are very similar to that of flows generated by a thermal gradient and a thermally responsive interface, except for the absence of density gradients.[32,33] The flow facilitates cargo assembly and stabilization at the storage.[34-37] Typical cargo particles are Polystyrene latex spheres denoted as PS$x$ with $x$ representing their diameter. The resulting assembly size, shape and dynamics strongly depends on the strength of the induced eo-flow. We qualitatively discriminate three characteristic situations. At low eo-flow, competition with outward electrophoretic motion of the cargo leads to the formation of single crystals beyond a void region.[35] Large eo-flow leads to the formation of local convection cells at the IEX which under swimming are loaded with many cargo particles.[28] In the present study, we are concerned with moderate flow conditions, under which the cargo assembles close to the IEX in a regular fashion and is not lifted up by convection.

In this work, we go beyond the previous works in two important points. First, we changed the irregular shaped IEX splinter to spherical IEX, which generates a symmetric electric field and eo-pump flow, thus allowing us to distinguish the roles played by each component. Second, we systematically varied the experimental boundary conditions, e.g., size and electro-kinetic mobility of cargo, size of IEX, cell geometry, surface charge of substrate, type and concentration of background salt. This allows us to propose a geometric model for the assembly. Moreover, it allows observing a robust characteristic cargo number dependent swimmer speed, which first increases stepwise, then shows an independence on cargo number. From our observations we can propose an active role of the cargo in breaking the symmetry of flows and hence in propulsion.

In what follows we will first present and then discuss the main results of this comprehensive experimental study. For details on the experimental procedures, additional results and videos, the interested reader is referred to the Supporting Information (SI).

**Results**

**From pump to swimmer**

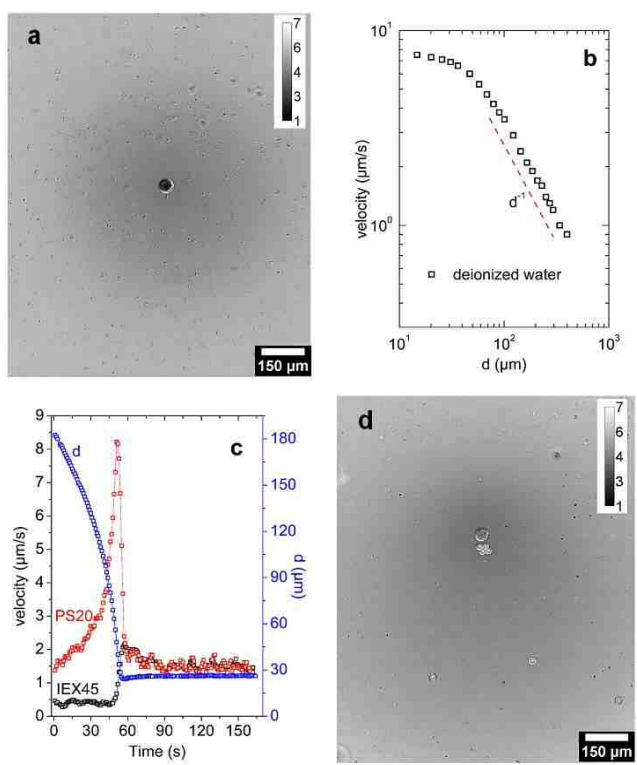

**Fig. 1:** (color online) a) pH gradient image of an IEX45 based pump with pH values indicated by the grey-scale. b) PS2 tracer velocity as a function of distance ($d$, measured from the projected surface of a fixed IEX45 to the center of tracer) in deionized water. The dashed line with a slope of -1 in this double logarithmic plot represents the theoretical expectation for a quasi-2 dimensional flow. c) Velocities (left axis) and corresponding change of the projected surface-to-center distance between IEX and cargo, $d$ (blue line, right axis) of an individual IEX45 (black line) and a single PS20 (red line). Shown is the approach, the locking at 53 s

and the final propulsion of the assembly. d) pH gradient image of a self-propelling swimmer formed of IEX67 and four PS31 with pH values indicated by the grey-scale.

Fig. 1a shows the stable symmetric pH gradient (~5 mins after IEX contacts with indicator solution, see SI) established by an immobile IEX45 based pump, which extends to a radial range $d$ of ~500 μm. The corresponding diffusio-electric field induces a converging eo-flow in the radial range $d$ of about 300 μm (Fig. 1b).[29-31] Due to the symmetry of flow, an individual IEX45 does not show directed motion by itself. Cargo particles in the range of eo-flow are advected towards the IEX (Fig. 1c). As one PS20 approaches, IEX45 starts moving at a projected surface-to-center distance between IEX45 and cargo of $d \approx 50$ μm; $d$ continuously decreases until at $t = 53$ s, it reaches a minimum and IEX and cargo become "locked" together. The complex keeps moving with IEX45 in the lead. This allows discriminating the front and the back of a swimmer. Swimming is accompanied by an asymmetric pH gradient shifted towards the back of the swimmer as shown in Fig. 1d.

**Assembly of modular micro-swimmers**

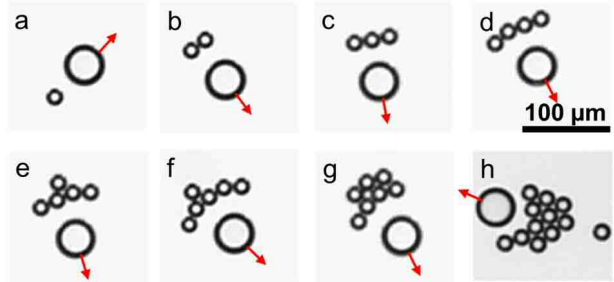

**Fig. 2:** (color online) Optical micrographs of linearly self-propelling swimmers formed of IEX45 and PS20: from a) to g), the total number of assembled cargos $N$ increases. For $N \geq 4$, a second row of cargo forms. h) For $N > N_{MAX}$, surplus cargo is ejected from the back of the swimmer. The directions of motion are indicated by the red arrows. The scale bar shown in d) applies for all images.

At low cargo density, swimmers typically form through one-by-one cargo uptake (videos 1, 2 and 3 in the SI). The typical assemblies of linearly self-propelling swimmers are shown in Figs. 2a-h. Cargos assemble at a well-defined assembly distance $d$ defined as the projected surface-to-center distance between IEX and the first-row cargo. Once the first row is filled with a maximum number of cargo, $N_{1,MAX}$, a second and further rows of cargo with decreasing member number form. Addition of further rows of cargo leads to more or less well ordered rafts of hexagonal close packing dictated by the spherical shape of the cargo. When the maximum cargo number, $N_{MAX}$, is reached, surplus cargos are ejected from the back and mark the trace taken by the swimmer (Figs. 2h and S1). Then, the total number of loaded cargo $N$ stays close to $N_{MAX}$ with some minor fluctuations depending on the details of the assembly shape and the presence or absence of small scale convection in the swimmer tail.

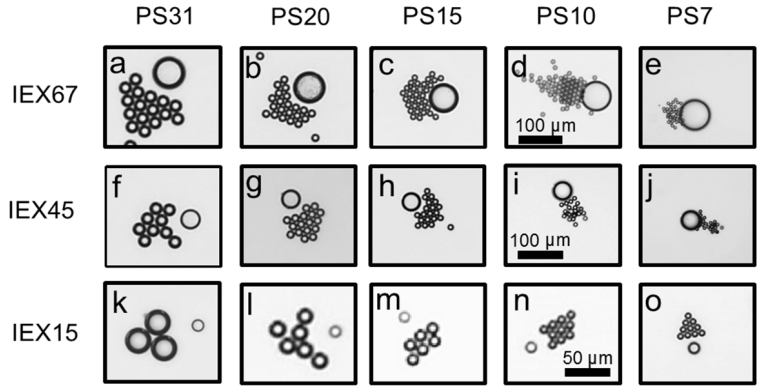

**Fig. 3:** Optical micrographs of modular micro-swimmers combined from different IEX and cargos. Note the different scale bar used for swimmers lead by IEX15.

**Table I**. Assembly parameters: cargo size dependence of $N_{MAX}$ and of the maximum opening angle of the first-row cargo $\alpha_{1,MAX}$ for differently sized IEX. Videos on IEX45 and cargo type as indicated can be found in the SI.

| Cargo | Video No. | $N_{MAX}$ @ IEX15 | $N_{MAX}$ @ IEX45 | $N_{MAX}$ @ IEX67 | $\alpha_{1,MAX}$ (°) @ IEX15 | $\alpha_{1,MAX}$ (°) @ IEX45 | $\alpha_{1,MAX}$ (°) @ IEX67 |
|---|---|---|---|---|---|---|---|
| PS31 | 3 | 3 | 8±2 | 16±2 | | 94.2±6.3 | 102.0±2.4 |
| PS20 | 1,4 | 5±2 | 15±2 | 19±2 | 35.4±2.8 | 118.1±9.2 | 105.5±7.5 |
| PS15 |  | 6±2 | 16±2 | 35±6 | 61.2±3.3 | 98.6±2.9 | 95.0±8.4 |
| PS10 | 4 | 9±2 | 17±4 | 49±8 | 62.8±6.0 | 96.9±9.7 | 74.5±2.0 |
| PS7 | 2 | 10±2 | 20±6 | 50±8 | 54.8±4.4 | 64.8±2.3 | 65.0±2.0 |

The appearance of swimmers at $N_{MAX}$ is striking in that all types of IEX transport a large number of cargos of much larger volume than their own (Fig. 3 and videos 1-3 and 5 in the SI). Typically for moderate eo-flow, cargos generally stay on the substrate to form a mono-layer raft in the wake of IEX. Only for the combinations of big IEX and very small cargo, no rafts form, and the cargo is partially loaded into a small-scale convection cell in the swimmer tail (Figs. 3e, 3i, 3j and video 2). For the special combination of IEX67 and PS10, cargos are in addition advected to form a second layer. Thus this combination has a larger $N_{MAX}$ (Fig. 3d). We quantify $N_{MAX}$ for all combinations of IEX and cargos, and compile the results in Table I. $N_{MAX}$ decreases with the size of cargo and increases with the size of IEX with slight fluctuations. Due to the layered ordering of cargo, there is a finite range of opening angle $\alpha$ enclosed by tangents from the center of IEX to the outer surface of cargo (defined as $\alpha_1$, if measured for the first-row cargo). $\alpha_1$ increases linearly with increasing number of first-row cargo $N_1$, but the maximum open angle $\alpha_{1,MAX}$ never exceeds 120° for single-row swimmers (see also Table I). Note that, at higher cargo density, a larger $N_{MAX}$ and a less well defined $\alpha_{1,MAX}$ are observed, because of the frequent cargo-upload and stronger cargo-cargo interaction. Another observation from Fig. 3 is a diminishing assembly distance from left to right or from bottom to top of the figure, corresponding to a smaller $d$ with decreasing size of cargo or increasing size of IEX (Fig. S2). We further note, that the assembly behavior at moderate flow is different to that observed at large flow conditions. The latter is realized using IEX splinters under otherwise identical conditions. There larger convection cells form at the IEX splinter and cargos are circulated within (Videos 5-7 in the SI).

**Velocity of modular micro-swimmers**

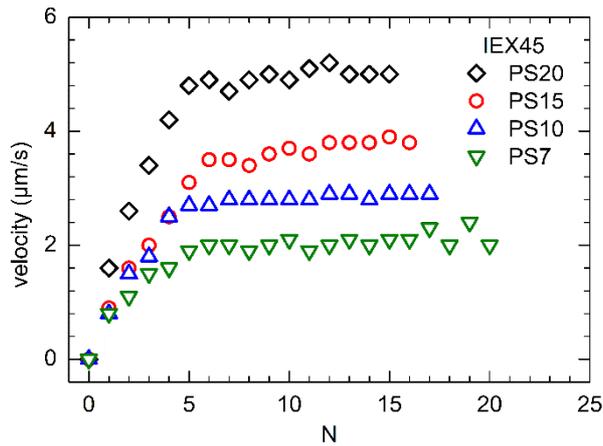

**Fig. 4:** (color online) Velocity of complexes lead by IEX45 versus number of assembled cargo $N$ of different sizes.

The most remarkable finding of the present study is shown in Fig. 4. With increasing cargo number, the swimmer velocity first increases, then saturates at a limiting velocity. We observe this characteristic dependence irrespective of experimental boundary conditions (Figs. 5, S3 and S4). At fixed number of cargo, the direction and velocity of an individual spherical IEX based swimmer vary only very little. They vary only during cargo-upload and in particular, if this is occurring from the front of a swimmer (Fig. S5 and videos 1-3). Therefore, we quantified the velocities only for conditions of one-by-one cargo-upload with individual upload events separated by at least 3 min. Each data point in Fig. 4 represents an average over 80-120 swimmers. A typical velocity distribution is shown in Fig. S6.

With increasing $N$, the velocity of swimmer increases in almost equal-sized steps. It saturates when $N$ = 5-6 and stays constant up to $N_{MAX}$. Fig. 4 further shows that there is a characteristic load number, at which the velocity changes its dependency on $N$. Incidentally, for spherical IEX, this number almost coincides with the maximum first-row loading capacity $N_{1,MAX}$ for most cargo sizes. Fig. 4 also shows a linear increase of the plateau velocity with the size of cargo. However, a systematic investigation on different combinations of IEX and cargos shows that the relation between swimmer speed and the size of cargo/IEX is more complicated (c.f. Fig. S3). Qualitatively, the fastest speed is observed at a size ratio of IEX-to-cargo of about 2.

**Influence of experimental boundary conditions**

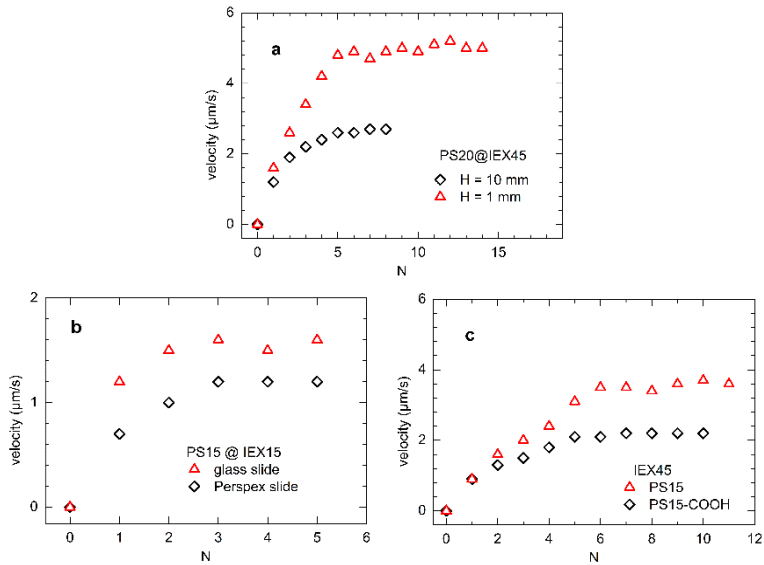

**Fig. 5:** (color online) a) Velocity of complexes lead by IEX45 versus number of PS20 at different cell heights. b) Velocity of complexes lead by IEX15 versus number of PS15 cargo on different substrates. The literature value of the surface potential for Perspex is -30 mV,[38] our glass slides have a surface potential of -110 mV under deionized condition (see SI). c) Velocity of complexes composed of IEX45 and cargo particles of PS15 and PS15-COOH. PS15-COOH has a lower electro-phoretic mobility but the same size and mass density compared with PS15.

While the general shape of the velocity-cargo number curve is very robust, the maximum speeds attainable vary systematically with experimental boundary conditions. Next, we investigate their influence in order to distinguish the roles played by each component in modular swimming. Motion of the complex depends on the overall cell geometry. Fig. 5a compares swimming speeds in cells of height $H = 1$ mm to speeds measured at $H = 10$ mm. For pumping at an immobile IEX, such an increase in $H$ was shown to cause a qualitative change of the distance dependence of the eo-pumping velocity which switches from $1/d$ to $1/d^{1.8}$ and a slight increase of the maximum tracer velocity originating from different distributions of pH gradient.[29] Here, the swimming velocity halves as the cell height is increased. At the same time, the assembly distance increases.

The eo-pump flow can be reduced by utilizing substrates of lower eo-mobility. We have shown that this works well *via* a reduced surface potential obtained by coating the substrate with a suitable cationic poly-electrolyte.[35] In Fig. 5b, we demonstrate the influence of a lowered electrostatic surface potential at otherwise constant conditions on the swimming behavior. We compare the swimming speeds of complexes formed of IEX15 and PS15 on uncoated glass slides ($\zeta = -110$ mV, see SI) and on Perspex slides ($\zeta = -30$ mV[38]). Fig. 5b shows that the swimming velocities on the Perspex substrate are systematically lower than those on the glass substrate.

Changing the cargo material leaves the eo-pump flow unaltered but may change the electro-phoretic (ep) mobility ($\mu_{ep}$, see experimental part in SI) and the mass density ($\rho$) of the cargo. In Fig. 5c we test the influence of cargo material on swimming speed using PS15 ($\mu_{ep} = 2.5 \times 10^{-8}$ m$^2$V$^{-1}$s$^{-1}$) and PS15-COOH ($\mu_{ep} = 2.1 \times 10^{-8}$ m$^2$V$^{-1}$s$^{-1}$). Here, the density of both cargo species is the same ($\rho = 1050$ kgm$^{-3}$). Also the assembly distance at IEX45 is identical for both PS cargos at $d = 14.7 \pm 3.4$ μm. Fig. 5c shows that the swimming speeds of complexes formed by cargos of lower $\mu_{ep}$ are significantly slower. A similar decrease of swimming speed of complexes formed from IEX45 and a silica cargo of lower $\mu_{ep}$ and larger density is shown in Fig. S7.

**Discussion**

The present investigation focuses on the assembly of spherical cargos in the wake of a mobile spherical ion exchange particle and the self-organization of directed motion of the formed complex under conditions of moderate solvent flows. This micro-scale example of the self-organized motion is noteworthy because none of the involved constituents is capable of directed motion by itself. Even more remarkable is the repeatability of the observed cargo assembly, cargo arrangement, trail formation and the directed motion with a defined velocity curve under changes of experimental boundary conditions.

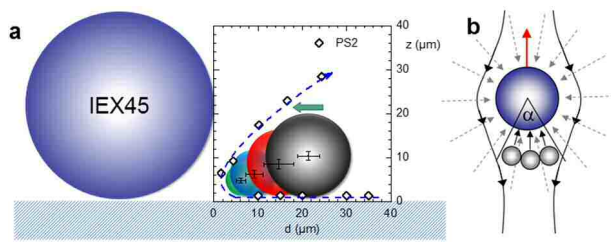

**Fig. 6:** (color online) Assembly during modular micro-swimming. a) To-scale sketch (side view) for differently sized cargo locked at their respective assembly distances *d* from the projected surface of IEX45. Open diamonds show the trajectory of the nearly buoyant PS2 tracing the solvent flow. The dashed line serves as a guide to the eye. Error-crosses denote the statistical uncertainty of cargo positions in *d*-direction and assume an uncertainty in *z*-direction of order 1 μm. The big green arrow denotes the electro-phoretic flow from the assembled cargos. The assembly distance increases with cargo size. b) Qualitative sketch in top view with three cargos assembled in the wake of an IEX bead. The outer edges of the outer cargos form a solid opening angle $\alpha$. The outer solid lines (black) denote the solvent streamlines in the reference frame of the IEX particle ("head wind"). This relative flow limits $\alpha$. The dashed grey arrows denote the symmetric eo-pump flow along the substrate surface pointing inward towards the IEX. The small solid arrows denote the superimposed additional electro-phoretic flow arising at the assembled cargos. The big red arrow denotes the lab frame propulsion direction of the complex.

We first discuss the observed assembly distance *d*, which increases with increasing size of cargo for any combination of IEX and cargo (Figs. 3 and S2). Fig. 6a shows a to-scale sketch (side view) of the assembly distances of cargo under moderate solvent flow. The sorting by size is inversed as compared to that observed under low flow conditions.[35] Here, the assembly distance appears to be limited by the dashed line denoting the solvent flow as traced by near buoyant PS2 and also found in simulations of solvent flow at a stationary IEX45[29]. Big massive cargos are advected towards the IEX but not lifted up by the convective flow. In vicinity of the IEX, the front and upper surfaces of the cargo experience a substantial upward and backward solvent flow, respectively. An influence of direct electro-static repulsion is unlikely due to the small Debye screening length $\kappa^{-1} \sim (10\text{-}50)$ nm as compared to $d \cong 6\text{-}42$ μm. A dominant influence of electro-phoretic flow on *d* can be ruled out from the observation of the same assembly distance for PS15 and PS15-COOH. These two types of cargo have identical size and density but significantly differ in their $\mu_{ep}$. We therefore propose the assembly distance at moderate flow to be determined mainly by the solvent flow geometry.

The idea of an advective limitation of *d* is further supported from the observation of the transition from a non-rotating to a rotating state for the asymmetric cargo (see video 4). Rotation about an axis perpendicular to the propulsion direction and parallel to the substrate can also be seen for Janus-type cargo particles.

Besides the well-defined assembly distance, the complex shows a regular assembly shape. The first incoming cargo assembles in the wake of the IEX and sets the complex into motion (in lab frame). More cargos are advected by eo-flow and take their positions next to the first, forming a single row. Under swimming condition, the solvent is moving against the swimmer resulting in a pronounced "head wind". Thus, cargo docking at the front drifts to the back of the complex (videos 1 and 2 in the SI). At small first-row opening angle, the radially inward solvent flow acting on the cargo can balance the solvent "head wind". Beyond a critical first-row opening angle, the motion of the solvent cannot be balanced any more, and cargo is pushed from the outer edge to the next row and back into the wake. This yields a maximum opening angle $\alpha_{1,MAX}$ for a single first row which never exceeds 120°. This situation is illustrated in top view in Fig. 6b, with representative flow lines shown in the reference frame of the IEX. Even when swimming at constant speed and direction (lab-frame based red arrow), the inward eo-pump flow prevails (grey dashed arrows) while the overall solvent has a relative velocity with flow lines engulfing the swimmer (solid large arrows). In addition, there is a superimposed electro-phoretic flow originating at the cargo surface (small solid arrows).

The true solvent flow is presumably even more complex. One example is shown in video 4 in the SI. It shows a wiggling motion of PS10 at IEX45 when getting in contact with the solvent "head wind" as the complex takes a right turn. This shows that the solvent may not follow simple trajectories and even more complicated trajectories of cargo particles may result. In addition, Fig. 6b falls short in considering any out of plane components of solvent flow as well as possible deviations from radial symmetry of eo-pump flow, which can be expected from the asymmetric pH gradient in Fig. 1d and the presence of cargos acting as geometric obstacle. Finally, in video 3 of the SI, one may notice systematic motion of particles far off the moving complex (reminiscent of a bow wave) which are not expected from the picture sketched in Fig. 6. However, we believe that Fig. 6 captures the main essentials of our qualitative geometric model under moderate flow conditions and suffices for a first discussion of the underlying physics.

The most striking observation made is the dependence of swimmer velocity on the number of cargo, which is robust against any changes of experimental boundary conditions and a characteristic feature of modular micro-swimming under moderate solvent flows (c.f. Figs. 4, 5, S3 and S4). In particular, the increase at $N < N_{1,MAX}$ is somewhat against intuition, because in the macro-world any additional load usually slows the transport via an increased friction. It is also contrary to observations on micro-swimmers formed from a single active component, for which the transport velocity also decreases with increasing number of cargo.[39-44]

For the present modular swimmers, the underlying mechanism for propulsion is a symmetry breaking of the originally radially symmetric inward solvent flow. This is first realized by the presence of assembled cargos, although a precise description of this contribution needs a full 3-dimensional simulation to capture its out of plane components and is thus beyond the scope of the present paper. Second, with cargo loaded in place, there is an additional ep-flow along its surface, which is inward directed and superimposed on the eo-pump flow generated on the substrate. This directly follows from the experiments shown in Fig. 5c. There, all the geometric boundary conditions are the same for the two different types of cargo: same cell height, same cargo size and density, same IEX and substrate, and same assembly distance. Yet, the complexes formed from the cargo of larger $\mu_{ep}$ show the systematically larger speeds, irrespective of $N$. Thus the additional ep-flow from the assembled cargos actively breaks the symmetry of solvent flow. Our observation of an active role of cargo is also well compatible with the characteristic dependence of swimming speed on cargo number. At low $N_1$, each added cargo adds another ep-flow field. Thus the swimming velocity increases stepwise. Note that the steps are first equal in size but somewhat smaller as $N$ approaches $N_{1,MAX}$, which we believe is an effect of the gradual change in ep-flow direction with increasing $\alpha$.

To explain the saturation of velocities within this picture we consider i) the constant ep-mobility of the cargo and ii) the $1/d$ decrease of the diffusio-electric field $E$[29] which yields iii) a $1/d$ dependence of ep-flow velocity $v_{ep} = \mu_{ep}E$ well compatible with the $1/d$ dependence required for converging 2-dimensional flows of incompressible solvent. This provides a stability criterion for the assembled rafts but at the same time leaves $v_{ep}$ constant in the first row. With a solvent "head wind" controlled $N_1=$ constant, the swimmer velocity then saturates. The limiting velocities are found to be largest for a size ratio of IEX-to-cargo of about 2. We note that within our simple model (Fig. 6a), only for this size ratio, the ep-flow tangential to the cargo surface can impinge at the IEX center.

A third type of symmetry breaking occurs only under already established swimming motion. Quantifying the effects of pH gradient symmetry breaking affords extensive additional experiments and is not addressed here. The importance of this contribution remains an open question.

A final interesting point is the stability of the swimming direction intimately connected to the issue of steering. Videos 1-5 show that the moving direction changes upon cargo uptake or rearrangement, but returns to straight propulsion as soon as the symmetry of cargo assembly is restored. This effect is most pronounced at low cargo numbers. Under conditions of large flow (observed in videos 6 and 7 but studied in more detail in ref. 28) the motion of complexes stays directed despite the rapid rearrangement of cargo loaded into its convection roll. Under conditions of moderate flow, symmetry may be broken permanently by using two cargos of different ep-mobility or different size (video 4). We also tested other cargo types and found that Pt-capped Janus particles in the presence of added $H_2O_2$ do not form stable cargo arrangements. Rather, we observe some local irregular motion of Janus particle in the wake of the IEX corresponding to an erratic motion for the complex of ill-defined and constantly changing direction. Conversely, any control of cargo arrangement by external fields, e.g., electric field for Janus particles[45-47] or magnetic fields for para-magnetic cargos[48,49] may in future be used to guide the swimmer to some destinations by remote control. In terms of scaling down to Brownian motion controlled region, modular swimmers, e.g., combination of IEX5 and PS2 are still self-propelling. The influence on directionality will be explored in future work.

Above, we have proposed a simple geometric picture which is qualitatively compatible with our observations. It further allows assigning different functionalities to different components of the modular swimmer by drawing macroscopic analogies. This is done in Fig. 7. The IEX exchanges stored $H^+$ for impurity cations and thus provides the necessary electrolyte gradients, It therefore acts as fuel reservoir or *storage*. The substrate provides the eo-flow. It acts as a *motor* setting the solvent in motion and as *clutch* coupling *storage* and *cargo* by flow. The *cargo* is obviously transported, but also acts as *gearing* providing different speeds for different *N* and as *rudder* implementing some steering through its assembly shape. While this is nice, we are, however, aware that we are still far off any rigorous quantitative modeling[24-27,50,51]. In fact we face a most complex interplay of different parameters, also shown in Fig. 7. Here, the experimentally controllable boundary conditions are labelled blue. They are: substrate ζ-potential ($\zeta_{substrate}$), ion exchange rate ($J^+$) and size of IEX ($\sigma_{IEX}$), cargo size ($a_{cargo}$) and ζ-potential ($\zeta_{cargo}$), type, concentrations and diffusion coefficients of background electrolytes ($z_i$, $n_i$ and $D_i$), dielectric constant ($\varepsilon$) and temperature ($T$) of solvent, and finally cell height ($H$). These parameters directly influence the diffusio-electric field ($E$), eo- and ep-flows, assembly distance $d$ and hovering height $z^*$ of cargo. The latter quantities are indicated by italics. They in turn combine to determine the self-assembly and swimming performance.

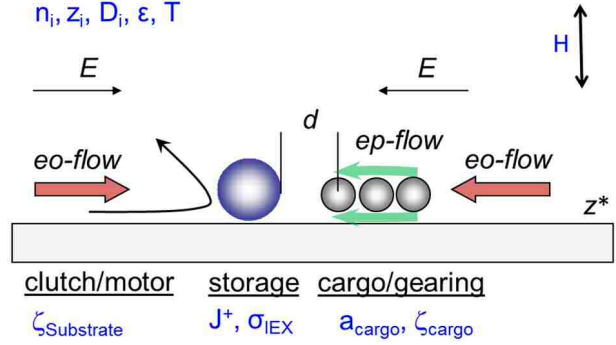

**Fig. 7:** (color online) Sketch of the situation in modular micro-swimming in a vertical cut along the propulsion direction with three rows of cargo. Experimentally adjustable quantities are marked in blue. They in turn influence quantities presumably controlling the self-assembly and swimming performance which are indicated by italics. Resulting functions are labelled in analogy to their macroscopic counter-parts and shown underlined. The outer large arrows mark the electro-osmotic solvent flow along the substrate. The thin black arrow represents a solvent flow line. The arrows next to the cargo mark the additional local electro-phoretically driven solvent flow (for details see text).

The relations between these controlling factors need to be solved theoretically step-by-step. We have measured the pH gradient generated by an immobile IEX, however, no theory can so far predict the $E$ field in this constant ion concentration solution. A full 3-dimensional simulation with all the hydrodynamics and boundary conditions considered should give consistent results with experimental observations including the assembly shape, the cargo size dependence of assembly distance and the typical velocity curve. At present, this constitutes a formidable challenge. The whole system is an inseparable entity including IEX and solvent, cargos and their surfaces, the responsive charged substrate, and ions, and the complexity of geometry under propulsion is much higher than that at rest and far exceeds that of a hot solid particle and a gas bubble[24,25] or a connected catalytic dimer[26,27]. Neglecting of the out of plane advection was sufficient for modelling the assembly under low flow.[35] However, it is not applicable for the present case of moderate solvent flow, under which the convective flow appears to induce a reversed assembly sequence of cargo. Up to now, we have modelled a stationary IEX-based eo-pump flow for the solvent in 3 dimensions, which gave qualitative agreement with experiment.[29] Proceeding to mobile pumps, additional cargo particles, advection of solutes at high Péclet number, symmetry breaking and propulsion of the assembled complexes make the modelling more demanding and computationally expensive. Results will be reported in future works.

**Conclusion**

The present paper introduces modular micro-swimming as a novel, versatile and effective experimental approach for directed cargo transport on the micro-scale. We systematically varied the experimentally accessible boundary conditions and determined the effects of these quantities on the observable self-organization and swimming performance (shape of complexes, assembly distance, loading capacity and cargo number dependent swimming speed). Based on our extensive experimental investigation, we developed a geometric model of self-assembly in modular swimming as an important intermediate step towards a full understanding of modular micro-swimming. Moreover, we have proposed possible ways for steering. We therefore hope to have stimulated other interesting experiments along the line of modular swimming. We further distinguished the roles played by each component in propelling which yielded important qualitative constraints for future modeling. Development of a full 3-dimensional model providing quantitative connections between experimentally accessible control parameters and observed assembly, motion and functionality of the swimmers remains an interesting challenge to theory.

**Acknowledgement**

We are pleased to thank H. Löwen, A. Ivlev, Chr. Holm, J. de Graaf, E. Oğuz and T. Speck for intense and fruitful discussions. We further thank Christopher Wittenberg for technical support in particle tracking programing and pH-measurements. Financial support of the DFG (SPP 1726 Project Pa459/18-1, 2) is gratefully acknowledged.

# Supporting Information

## Assembly and speed in ion exchange based modular phoretic micro-swimmers

Ran Niu*, Denis Botin, Julian Weber, Alexander Reinmüller, Thomas Palberg

Institut für Physik, Johannes Gutenberg-Universtät Mainz, Staudingerweg 7, 55128 Mainz, Germany

**Methods and Materials**

The main part of this study was carried out using micro-gel based cationic IEX spheres of diameters 20-100 μm (CGC50×8, Purolite Ltd, UK). These were sorted manually into diameter classes of 45±1 μm and 67±1 μm, denoted as IEX45 and IEX67, respectively. Size determination was performed by independent microscopy experiments on fixed beads by scanning the focal plane position.[S1] Smaller sized cationic IEX spheres of diameter 15.3±0.3 μm (CK10S, Mitsubishi Chemical Corporation, Japan, labelled as IEX15) and counter ions $Na^+$ were also used. To study the influence of shape (and enhanced exchange activity) we further used irregular shaped IEX splinters (lateral extension < 100 μm) which were obtained by crushing 1-2 mm IEX spheres (Amberlite K306, Roth GmbH, Germany) in a mortar. All resins were washed with 20 wt % hydrochloride acid solution several times to exchange all counter ions into $H^+$. Then they were rinsed with doubly deionized, degassed water several times until solution pH reached ~ 7 and gently dried at 80 ºC for 2 h before use.

In most of the experiments, the model cargo particles are commercial, negatively charged polystyrene (PS) spheres (MicroParticles GmbH, Germany). Most are stabilized by sulfate surface groups of different diameters as determined by the manufacturer using electron microscope (lab codes PS2 to PS31). We also used carboxylate modified PS of diameter 15 μm (PS15-COOH) and bare silica particles with diameter of 20 μm (Si20, AkzoNobel, Sweden). Electro-phoretic mobilities ($\mu_{ep}$) were determined by micro-electrophoresis in a home-build Perspex cell (10 mm×10 mm) based on the construction originally introduced by Uzgiris[S2]. An alternating square wave electric field ($f$ = 0.5 Hz; $E$ = ± 40 V/cm) was applied between two platinum electrodes mounted vertically in the cell center far off the cell walls to avoid influence of electro-osmosis. The cell was mounted on a micro-electrophoresis instrument (Mark II, Rank Bros. Bottisham, Cambridge UK) providing ultra-microscopic illumination and observed with a consumer DSLR (D800, Nikon, Japan) equipped with macro-lens. Using an exposure time of 3 s, the response of individual spheres to gravity and the field translates to zigzagged trajectories readily analyzed for the electro-phoretic mobility $\mu_{ep}$. Data were averaged over at least $5\times10^2$ particles for each species. Mobilities and other cargo particle data are summarized in Table SI.

**Table SI:** Parameters of cargo particles used in this work.

| Batch No. | Lab code | Diameter 2a (μm) | $\mu_{ep}$ ($10^{-8}$ m$^2$ V$^{-1}$ s$^{-1}$) |
|---|---|---|---|
| PS/Q-F-B1184 | PS31 | 31.1±0.3 | 2.7±0.3 |
| PS/Q-F-L2457 | PS20 | 19.7±0.2 | 2.6±0.2 |
| PS/Q-F-L1488 | PS15 | 15.2±0.1 | 2.5±0.2 |
| PS/Q-F-B1278 | PS10 | 10.7±0.1 | 2.5±0.3 |
| PS/Q-F-L771 | PS7 | 7.6±0.1 | 2.6±0.35 |
| PS/Q-F-B1203 | PS4 | 4.1±0.1 | 2.7±0.2 |
| PS/Q-F-L2090 | PS2 | 1.7±0.1 | 2.0±0.2 |
| PS-COOH BS860 | PS15-COOH | 15.7±0.2 | 2.1±0.2 |
| Kromasil100Å | Si20 | 20.9±5.2 | 2.3±0.3 |

The cell for swimming experiments was constructed from circular Perspex rings with diameter of $D$ = 20 mm attached to a microscopy slide by hydrolytically inert epoxy glue (UHU plus sofortfest, UHU GmbH, Germany) and dried for 24 h before use. Standard ring height was $H$ = 1 mm. In addition, rings with height $H$ = 10 mm were used for the flow geometry dependent measurements. Commercial soda lime glass slides of hydrolytic class 3 (VWR International, Germany) served as substrates. These were washed with 1% alkaline solution (Hellmanex®III, Hellma Analytics) under sonication for 30 min, then rinsed with tap water and subsequently washed with doubly distilled water for several times. Substrate surface potentials were determined using Doppler velocimetry[S3] employing a cell with exchangeable side walls. From the electro-osmotic mobility of the standard glass slides under thoroughly deionized conditions their ζ-potential was calculated using the standard electro-kinetic model[S4] as ζ = -110 mV. In addition, cells built using Perspex slides of surface potential $\zeta$ = -30 mV$^{S5}$ were used for the substrate dependent measurements. In a typical experiment, a few IEX spheres or splinters were placed inside the cell, followed by injection of 0.4 mL of colloidal particle suspension. Then the cell was quickly covered by another glass slide to avoid contamination by dust.

Samples were observed typically at 5× or 10× magnifications using an inverted scientific microscope (DMIRBE, Leica, Germany) equipped with a standard video camera. Colloidal particles quickly settle to the bottom of the cell, where they form a very dilute sedimentation layer with 4−12 particles in the field of view (typical size of 1086×814 µm$^2$). Videos were recorded at a frame rate of 1 Hz and analyzed using a self-written Python script. IEX particle appeared as a circle in the image, which can be mathematically expressed as: $(x-x_{center})^2+(y-y_{center})^2 = r^2$. Here $(x_{center}, y_{center})$ is the center of the circle, and $r$ is the radius. Hough transform implemented in the OpenCV function HoughCircle was applied to extract the edge from image gradient information. Then the edge was imported into above equation to extract the position of particles. Subsequently, from the positions of IEX particles at different frames, the trajectories and velocities of the formed modular micro-swimmers were calculated. Each velocity data is the average over 80−120 individual complexes. Cargo particles carried by spherical IEX were detected at the same time thus the surface-to-center distance ($d$) between IEX and the first-row cargo was also calculated and averaged for each cargo size over at least 150 particles.

The pH gradient within a fixed IEX based pump and a self-propelling swimmer was measured using 1:3 (volume ratio) mixture of universal indicators (pH 0-5 and pH 4-10, Sigma Aldrich, Inc.). RGB color-images were taken using a consumer DSLR (D700, Nikon, Japan) mounted on an inverted scientific microscope (DMIRBE, Leica, Germany). From the monotonic decrease of the color ratio of blue-to-red with increasing pH, we obtain a calibration curve at different fixed pH. Applying the calibration curve to the image of samples with small amounts of indicator solution, we can measure the pH value at each pixel around IEX. Images shown code the pH as B&W-intensity using 9 pixel gliding averages.

The solvent flow lines generated in the vicinity of an IEX bead are reconstructed from the trajectories of small PS2 (1.7 µm in diameter) particles. We placed the focus plane of the microscope at different heights and note the distances, at which images of PS2 appear sharp. As the particles are nearly buoyant, they are carried along with the upward/backward flow of the solvent, thus the trajectory of PS2 closely resembles the true solvent flow line.

**Videos of modular micro-swimmers:**

Video 1: **(Video 1.avi)** Modular micro-swimmer composed of IEX45 and PS20. One-by-one cargo uptake. Straight motion except for cargo re-arrangement. Speed increases for $N < N_{1,MAX}$. (Image size 1086 × 814 μm²; 6 × real time speed).

Video 2: **(Video 2.avi)** Modular micro-swimmer composed of IEX45 and PS7. The first row of PS7 (close to IEX45) appears to be locked at a stable position. Some additional cargo is loaded into a small-scale convection roll trailing behind. (Image size 450 × 337 μm²; 6 × real time speed).

Video 3: **(Video 3.avi)** Modular micro-swimmer composed of IEX45 and PS31 during successive cargo upload. Note the slow velocity apparent even at 10fps. At t = 1:28 a PS31 bumps into the four cargo swimmer initiating a drastic transient decrease of velocity and a near 45° turn of the propulsion direction upon cargo rearrangement. Note the apparently swimmer-avoiding motion of the cargo particles to the front of the swimmer, which is also seen for the PS7 initially at the lower middle of the image. (Image size 1086 × 814 μm2; 10 × real time speed).

Video 4: **(Video 4.avi)** Modular micro-swimmer initially composed of IEX45 and PS10. A PS20 approaches from the side and is filed at larger assembly distance. Due to asymmetric propulsion from the differently sized cargos the swimmer takes a right turn. PS10 is nearly ejected by the flow from motion but manages to get back and files between PS20 and IEX45. A bent triple cluster of PS10 approaches with fixed orientation but starts rotating about its long axis upon assembly. (Image size 450 × 337 μm²; 10 × real time speed).

Video 5: **(Video 5.avi)** Modular micro-swimmer composed of IEX splinter and PS15. The splinter rotates after cargo uptake with unchanged direction of propulsion. The swimmer changes direction upon cargo rearrangement. (Image size 1150 × 860 μm²; 10 × real time speed).

Video 6: **(Video 6.avi)** Modular micro-swimmers composed of PS7 and cationic IEX splinters. The larger one is of triangular shape and side length of ca. 150 μm; the smaller one is of compact, slightly elliptical shape with long axis of ca. 110 μm. The larger swimmer has cargo loaded mainly into a single large convection roll. The smaller swimmer has cargo in an ordered raft and a similarly sized convection roll. In both cases, surplus cargo is expelled at the back and marks the swimmer trace by a comet-like trail. (Image size 1150 × 860 μm²; 10 × real time speed).

Video 7: **(Video 7.avi)** Modular micro-swimmer composed of PS7 and elongated cationic IEX splinter of ca. 100 μm length. The swimmer orients its smooth long side to the front. A large convection cell in the back takes up cargo. Surplus cargo is spilled out to the back forming a comet-like trail. A smaller convection cell is also seen at the swimmer front lifting cargo over the swimmer. (Image size 450×340 μm²; 5 × real time speed).

**Additional Data and Results**

**Table SII:** The maximum number of first-row cargo $N_{1,MAX}$ for different combinations of IEX and cargo. From $N_{1,MAX}$, the maximum opening angle of first-row cargo $\alpha_{1,MAX}$ is calculated geometrically as $\alpha_{1,MAX} = 2 \times N_{1,MAX} \times \arcsin(a_{cargo}/(d+a_{IEX}))$ (see Table I, main text). Here, $d$ is the assembly distance, $a_{IEX}$ and $a_{cargo}$ are the radii of IEX and cargo.

| Cargo | $N_{1,MAX}$ @ IEX15 | $N_{1,MAX}$ @ IEX45 | $N_{1,MAX}$ @ IEX67 |
|---|---|---|---|
| PS31 |  | 3 | 4 |
| PS20 | 2 | 5 | 5 |
| PS15 | 3 | 5 | 5 |
| PS10 | 4 | 5 | 5 |
| PS7 | 4 | 5 | 5 |

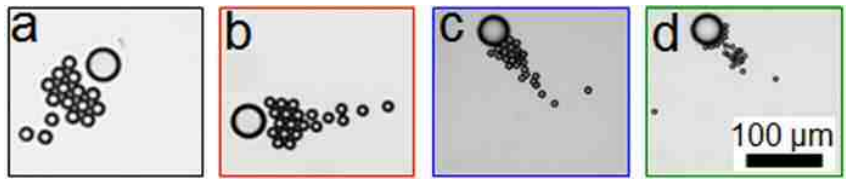

**Fig. S1.** Surplus cargos are ejected from the back of swimmers lead by IEX45 at $N > N_{MAX}$. From left to right, the cargo particles are PS20, PS15, PS10 and PS7.

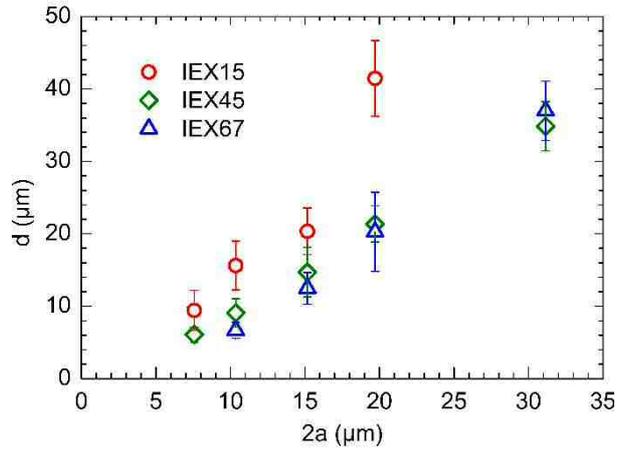

**Fig. S2.** Assembly distance of different PS cargo at IEX of different size. Here, $d$ was determined as an average over all first-row cargo particles at $N \leq N_{1,\text{MAX}}$. A systematic increase of $d$ with increasing size of cargo and decreasing size of IEX is observed.

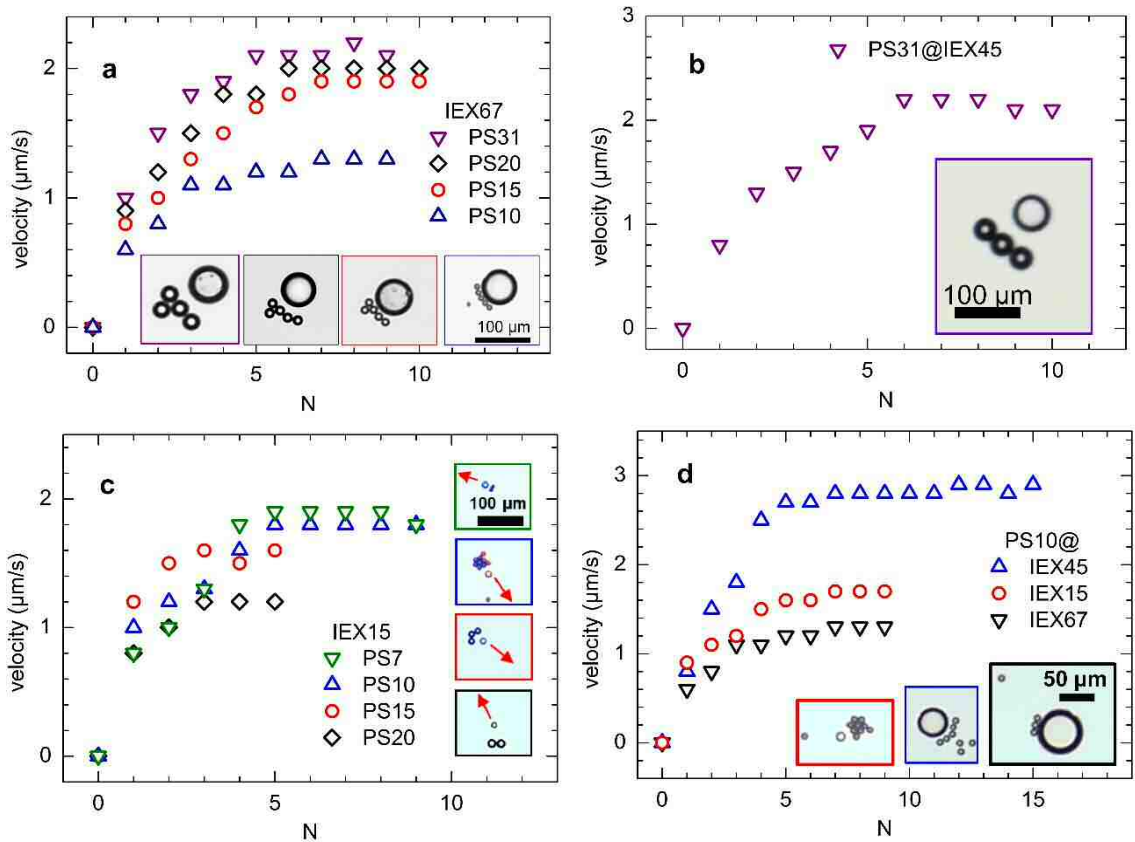

**Fig. S3.** Velocities of complexes formed from different combinations of IEX and PS cargo particles versus cargo number $N$. a) Variation of PS size at IEX67. Insets: corresponding micrographs (from left to right, cargo size decreases). b) PS31 at IEX45 versus cargo number. Inset: corresponding micrograph. c) Variation of PS size at IEX15. Note the inversed sequence of maximum velocities as compared to a). Insets: corresponding micrographs (from top to bottom, cargo size increases; red arrows indicate the directions of motion) d) Variation of IEX size using PS10. Insets: corresponding micrographs (from left to right, IEX size increases).

For different combinations of IEX and cargo, the swimming velocity versus cargo number curves show the same characteristic shape. Qualitatively, the limiting swimmer velocity is highest when the cargo size is about half the size of IEX. However, no simple quantitative relation is found.

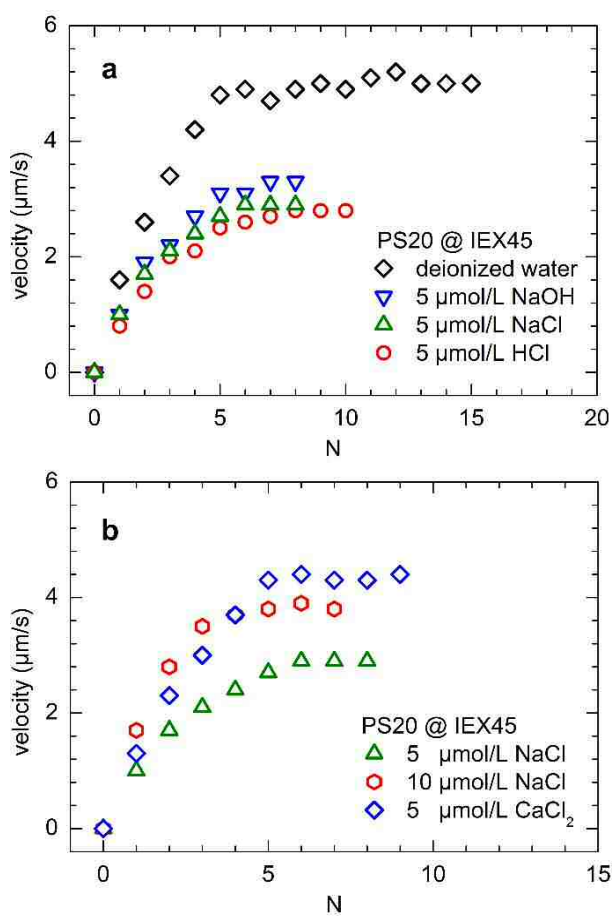

**Fig. S4.** (a) Velocity of complexes lead by IEX45 versus number of PS20 in deionized water and 5 μmol/L of different electrolytes. (b) Velocity of complexes lead by IEX45 versus number of PS20 at different salt concentrations.

Very similar, if slightly more rounded curves are also observed in low concentration electrolyte solutions. In Fig. S4a, adding 5 μmol/L of electrolyte, the velocity of complex decreases compared with that in deionized water. In Fig. S4b, adding 10 μmol/L of NaCl, the velocity of complex increases as compared with 5 μmol/L of NaCl. More interestingly, the velocity curve at 5 µmol/L of CaCl$_2$ and 10 µmol/L of NaCl, which have equivalent ionic strength, almost overlap confirming the origin of flow from ion exchange.

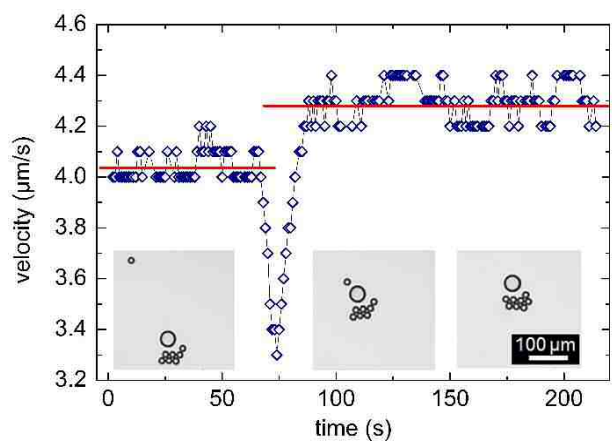

**Fig. S5.** Velocity of a swimmer composed of IEX45 and PS20 during cargo uptake. Red lines indicate the average velocities and are guides to the eye. Insets show micrographs taken during approaching, docking at the IEX front and after filing into the existing raft, increasing $N$ from 7 to 8. Note the pronounced dip in velocity during cargo-uptake from the front.

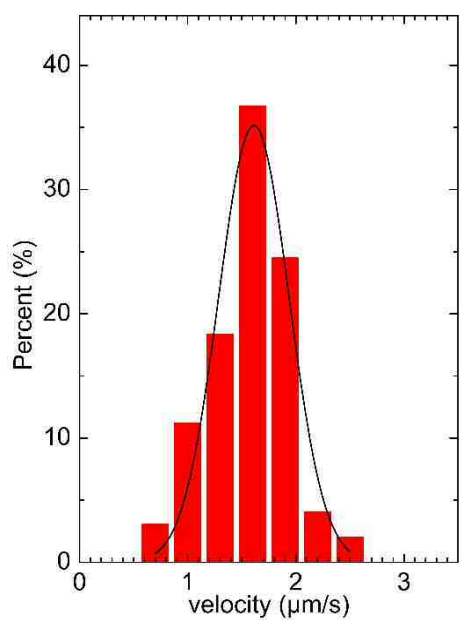

**Fig. S6.** Velocity distribution of complexes formed by IEX45 and two PS10. The distribution is well described by a Gaussian function with an average velocity of 1.6 µm/s and a standard deviation of 0.5 µm/s.

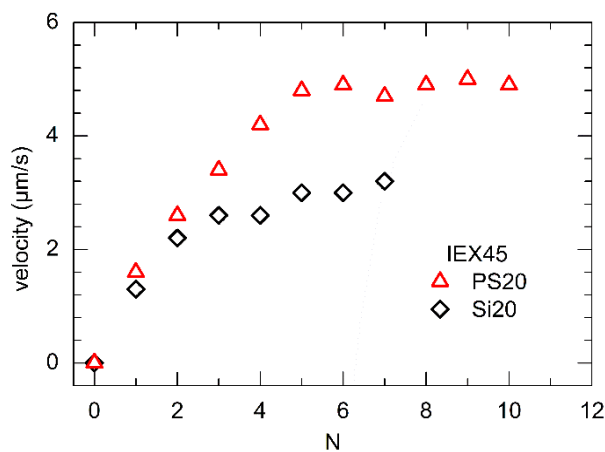

**Fig. S7.** Velocity of complexes composed of IEX45 and different cargo particles: comparison of PS20 ($\mu_{ep}$ = 2.6×10$^{-8}$ m$^2$V$^{-1}$s$^{-1}$, $\rho$ = 1050 kgm$^{-3}$) and Si20 ($\mu_{ep}$ = 1.9×10$^{-8}$ m$^2$V$^{-1}$s$^{-1}$, $\rho$ = 2540 kgm$^{-3}$). Complexes formed by IEX45 and PS20 show faster limiting velocities and a larger assembly distance of $d$ = 21.3 μm, and Si20 has a smaller assembly distance of $d$ = 17.5 μm.

Table of Content

# Assembly and speed in ion exchange based modular phoretic micro-swimmers

**Ran Niu\*, Denis Botin, Julian Weber, Alexander Reinmüller, Thomas Palberg**

Institut für Physik, Johannes Gutenberg-Universtät Mainz, Staudingerweg 7, 55128 Mainz, Germany

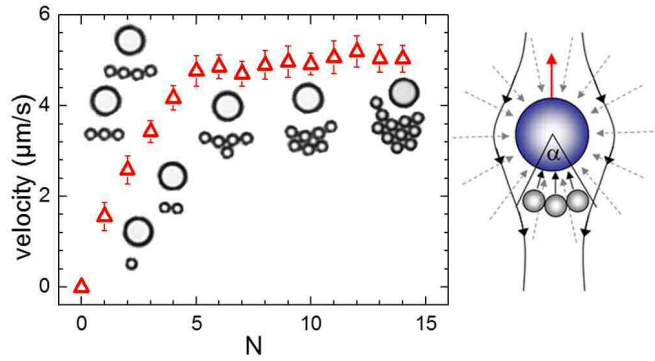

28